\documentclass[aps,onecolumn,nofootinbib,superscriptaddress,authoryear]{revtex4-2}

\usepackage[utf8]{inputenc}
\usepackage{amsmath,amssymb}
\usepackage{graphicx}
\usepackage{hyperref}
\usepackage{xcolor}
\usepackage{booktabs}
\usepackage{comment}
\usepackage{xspace}
\usepackage{aas_macros} 
\def\AK{{\tt AthenaK }}

\newcommand{\sobo}{\textrm{S$_1$B$_1$}\xspace}
\newcommand{\sobt}{\textrm{S$_1$B$_2$}\xspace}
\newcommand{\stbo}{\textrm{S$_2$B$_1$}\xspace}
\newcommand{\stbt}{\textrm{S$_2$B$_2$}\xspace}
\newcommand{\sthbo}{\textrm{S$_3$B$_1$}\xspace}
\newcommand{\sthbt}{\textrm{S$_3$B$_2$}\xspace}

\newcommand{\fobo}{\textrm{F$_1$B$_1$}\xspace}
\newcommand{\fobt}{\textrm{F$_1$B$_2$}\xspace}
\newcommand{\ftbo}{\textrm{F$_2$B$_1$}\xspace}
\newcommand{\ftbt}{\textrm{F$_2$B$_2$}\xspace}
\newcommand{\alfven}{Alfv\'en }
\def\Le{{\rm{Le}}}

\begin{document}

\title{Interior Magnetic Fields in Magnetars and Radio Pulsars}

\author{Raj Kishor Joshi}
\affiliation{Nicolaus Copernicus Astronomical Center, Polish Academy of Sciences, Bartycka 18, 00-716, Warsaw, Poland}
\author{Brynmor Haskell}
\affiliation{Department of Physics, University of Milan, Via Celoria 16, 20133 Milano, Italy}
\affiliation{INFN, Sezione di Milano, Via Celoria 16, 20133 Milano, Italy}
\affiliation{Nicolaus Copernicus Astronomical Center, Polish Academy of Sciences, Bartycka 18, 00-716, Warsaw, Poland}
\author{William Cook}
\affiliation{Theoretisch-Physikalisches Institut, Friedrich-Schiller-Universit{\"a}t Jena, 07743, Jena, Germany}
\author{Sebastiano Bernuzzi}
\affiliation{Theoretisch-Physikalisches Institut, Friedrich-Schiller-Universit{\"a}t Jena, 07743, Jena, Germany}

\date{\today}

\begin{abstract}
Magnetic fields are fundamental for neutron star physics and play a central role in powering the extreme phenomenology of magnetars, including Soft Gamma Repeaters and Anomalous X-ray Pulsars. However, the structure and stability of their internal magnetic fields remain largely unconstrained, as they cannot be directly probed by electromagnetic observations. Using 3D general-relativistic magnetohydrodynamics simulations across a range of rotation rates and magnetic field strengths, we identify two distinct evolutionary regimes leading towards dynamically stable magnetic configurations.
In rapidly rotating stars, the \alfven crossing timescale exceeds the rotation period, allowing differential winding to amplify a strong toroidal magnetic component before the onset of instabilities, leading to long-lived, stable configurations. In contrast, in magnetically dominated stars, instabilities in the poloidal field drive rapid field decay, leaving only a comparatively weak toroidal component.
These results imply that the internal magnetic structure of neutron stars depends sensitively on their rotational state: rotation-dominated stars like radio pulsars develop strong toroidal fields, while magnetars are characterized by predominantly poloidal configurations. Our findings therefore show that a neutron star's rotational history shapes its internal magnetic structure, providing a unifying physical picture that connects the observed diversity of neutron star classes to their hidden field configurations.
\end{abstract}

\maketitle

With surface strengths that can exceed $10^{15}$ G, magnetic fields play a central role in shaping neutron star (NS) dynamics and observables, yet their internal and near-surface topology remains poorly constrained. Most observational inferences, primarily from pulsar timing, probe only the large-scale field structure near the light cylinder, far from the star, where a dipolar description is generally adequate. In contrast, recent NICER observations reveal a substantially more complex magnetic geometry close to the stellar surface, with clear evidence for strong multipolar structure \cite{Bilous:2019,Riley:2021pdl,Salmi:2024}.

From a theoretical perspective, simple magnetic configurations are generically unstable: both laboratory plasma confinement experiments and astrophysical magnetohydrodynamics (MHD) studies show that long-lived equilibria require mixed poloidal--toroidal fields \cite{Braithwaite:2005md,Braithwaite:2009,Ciolfi:2012en,Lasky:2012ju,Sur:2021awe,Sur:2020hwn,Cook:2025zzy}, but the processes that determine their topology, relative strength, and long-term evolution are still debated.
Producing a stable magnetised NS configuration has remained an elusive goal. While the hydromagnetic equilibrium equations can be solved to construct magnetic NS equilibria both in Newtonian gravity and in General Relativity (GR), including physical effects such as stratification, superconductivity and the solid crust, the resulting configurations are not guaranteed to be dynamically stable \cite{Haskell:2007bh,Lander:2012a,Ciolfi:2013dta,Gusakov:2017uam,Sur:2021PASA...38...43S,Fujisawa:2023}. Dynamical simulations are therefore crucial both as direct models and as inputs for secular processes such as Hall drift and ambipolar diffusion \cite{Igoshev:2026ApJ..1000..291I}.

Numerical simulations in GR have confirmed that purely poloidal or purely toroidal configurations develop Tayler instabilities \cite{Markey:1973a,Tayler:1973a} on timescales comparable to the \alfven crossing time and drive the system toward mixed poloidal--toroidal configurations~\cite{Lasky:2011un,Kiuchi:2011yt,Ciolfi:2012en,Sur:2021awe,Cook:2025zzy}. Motivated by these stability results, NS magnetic fields have therefore often been modelled as twisted-torus configurations \cite{Lander:2009ib, Mastrano:2011, Lander:2012a, Lasky:2013}, with the further assumption that the internal toroidal component may greatly exceed the exterior poloidal field~\cite{Ciolfi:2013dta}. However, self-consistent simulations have struggled to realise such strong toroidally dominated equilibria. In particular, only a handful of studies have explored magnetic-field evolution in rotating systems~\cite{Cheong:2024stz, Pinas:2025bpq, Tsokaros:2021pkh}, and none, to our knowledge, in the low Lehnert number regime --- that is, the regime in which the rotation period is shorter than the \alfven crossing timescale. This regime is directly relevant for most of the pulsar population.

We perform a suite of 3D long-term general-relativistic magnetohydrodynamic (GRMHD) simulations with the code \AK (See Supplementary Material), and dynamically construct the first twisted-torus configurations in rapidly rotating NSs. These configurations are stable over 10--15 cumulative \alfven timescales, with no evidence of appreciable magnetic field decay. Our results show that the competition between rotation and magnetic stresses separates NSs into two distinct regimes with qualitatively different internal magnetic structures. We determine a general criterion separating the two regimes, demonstrating that magnetars are likely to have quasi-poloidal fields, while most pulsars may harbour strong internal toroidal components up to an order of magnitude stronger than their exterior dipole field.

We consider both slowly and rapidly rotating models ($\rm S$ and $\rm F$ series, respectively) at magnetic field strength $B_1 \approx 10^{16}$\,G, to map how increasing rotation rate alters the field evolution. These series of simulations are complemented by a $\rm B_2$ series, in which the field strength is either increased or decreased in each simulation in order to directly disentangle the competing \alfven and rotation timescales (see below and Supplementary Material). The relative impact of rotation on the magnetic field evolution can be quantified by the Lehnert number $\Le \equiv P/\tau_{\rm A}$ that measures the ratio of the rotational period $P$ to the \alfven crossing timescale $\tau_{\rm A}$. 

\begin{figure*}[t]
    \centering
    \includegraphics[width=0.95\textwidth]{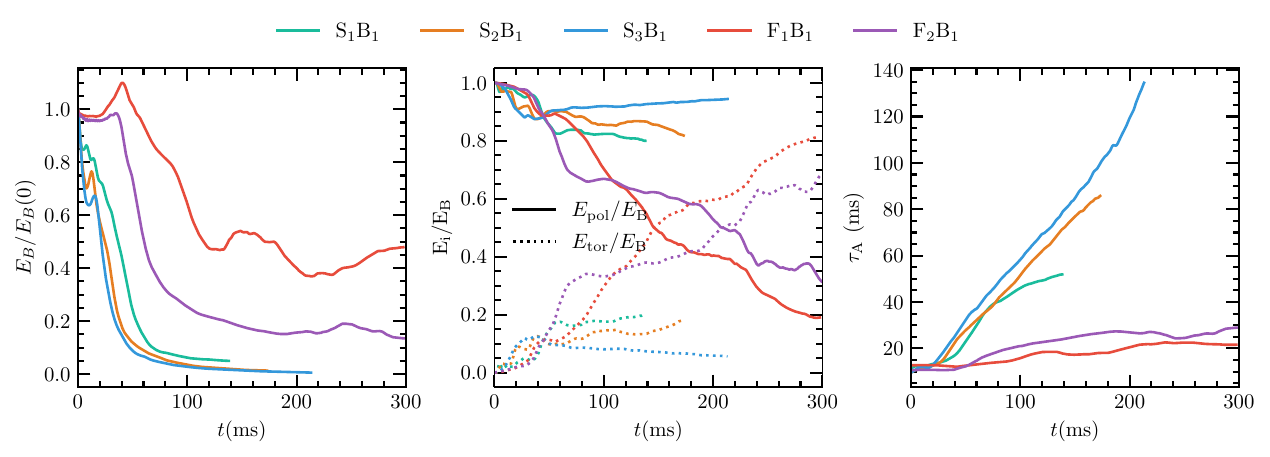}
    \includegraphics[width=0.95\textwidth]{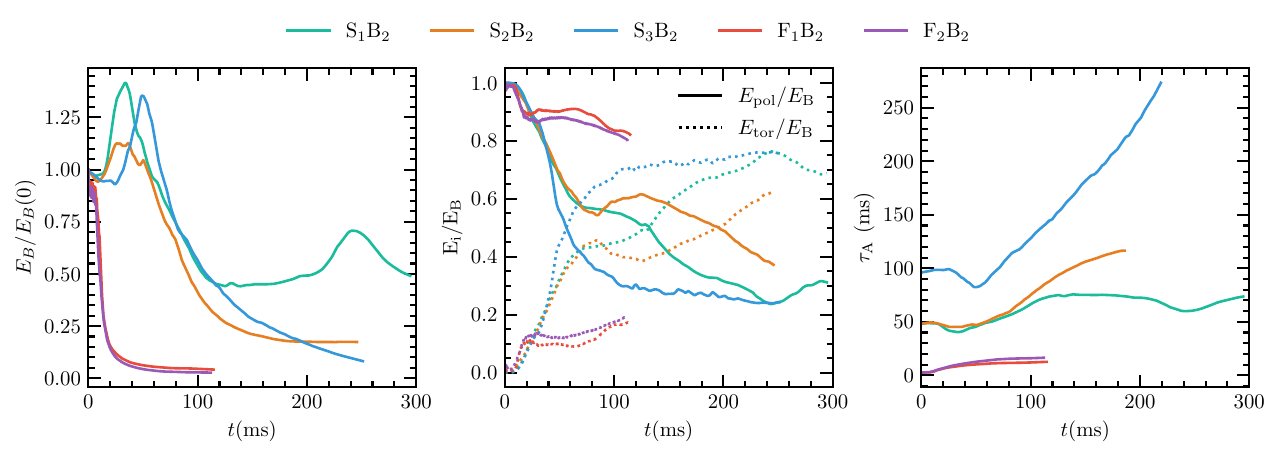}
    \caption{Evolution of total magnetic energy normalised by its initial value (left); toroidal and poloidal energy fractions (middle), and \alfven crossing time $\tau_{\rm A}$ (right) for fast-rotating F models (rotation period $P = 1.63$--$2.39$\,ms) and slow-rotating S models ($P = 10.84-50.0$\,ms) at field strengths $\rm B_1 \approx 10^{16}$\,G (top) and $\rm B_2$ (bottom).}
    \label{fig:energy_b1}
\end{figure*}

We begin by examining the $\rm B_1$ series, for which all models share an initial \alfven crossing timescale $\tau_{\rm A}$ exceeding ${\sim}10$~ms. The evolution of the magnetic energy, the toroidal and poloidal energy fractions, and the \alfven crossing time are shown in the top panel of Fig.~\ref{fig:energy_b1}. Consider the slowly rotating model $\rm S_1B_1$  characterized by $\Le=0.88$. In this case $\tau_{\rm A}\approx P$ and the run exhibits rapid magnetic energy loss due to reconfigurations caused by varicose and kink modes of the Tayler instability before the rotation can influence magnetic field evolution. The toroidal energy saturates at ${\sim} 20\%$ of the total. Nearly $90$\% of the total magnetic energy is lost by $100$~ms, and the associated field weakening drives a progressive increase in \alfven time, as shown in the rightmost panel. This behaviour is consistent with that observed in previous studies of non-rotating and slowly rotating simulations. Models \stbo ($\Le = 2.08$) and \sthbo ($\Le = 5.08$) follow the same
trend, with progressively weaker toroidal components and faster
magnetic energy loss as $\Le$ increases further above unity. 

For fast-rotating models \fobo ($P = 1.63$\,ms; $\Le=0.12$) and \ftbo ($P = 2.39$\,ms; $\Le=0.23$), the rotation period is shorter than the \alfven timescale and rotation delays the onset of magnetic energy loss. Model \fobo exhibits a transient growth in magnetic energy, and a comparison with \ftbo shows that faster rotation enhances the toroidal field. In model \ftbo the toroidal energy grows to approximately twice the poloidal energy by the end of the simulation, whereas in model \fobo this ratio reaches roughly four, indicating a substantially stronger toroidal component and helps sustain a larger fraction of the total magnetic energy.
As the faster rotation suppresses magnetic energy loss, the \alfven
timescale evolves more slowly, allowing these simulations
to accumulate significantly more cumulative \alfven crossing periods ($T_A$) within the
same coordinate time interval compared to the slowly rotating models.

The role of rotation and characteristic timescales is further highlighted by the ${\rm B_2}$ series of simulations. For the $\rm S$ models, the magnetic field strength in the $\rm B_2$
series is reduced to increase the \alfven timescale so that it is similar to the rotation period, whereas for the $\rm F$ models it is strengthened to decrease the \alfven timescale and bring it closer to the shorter rotation period. 
The evolution of the magnetic energies for these cases is shown in the bottom panels of Fig.~\ref{fig:energy_b1}. The longer \alfven time in model \sobt delays the onset
of the instability, opening a window for rotation to influence the field before significant losses occur. 
Consequently, model \sobt ($\Le=0.2$) shows a qualitatively different behaviour compared to \sobo ($\Le=0.88$); its evolution resembles that of models \fobo and \ftbo, which are dominated by rotation. In this case, transient magnetic energy growth is observed up to ${\sim} 50$~ms, reducing the \alfven timescale and eventually triggering the previously suppressed instability. The latter is followed by a phase of magnetic energy loss. As the system shifts toward a rotation-dominated regime, the toroidal component strengthens, reaching up to ${\sim} 60\%$ of the total magnetic energy.\\

Magnetically dominated behaviour analogous to models \sobo, \stbo, and \sthbo is recovered in the fast-rotating models when the field strength is increased. For model \fobt ($\Le=0.64$), the initial \alfven time is $2.53$~ms, close to the rotation period of $1.63$~ms. As a result, this model exhibits magnetically dominated behaviour similar to \sobo. The toroidal magnetic energy reaches only about $20$\% of the total by the end of the simulation. The similar behaviour of the slow and fast rotating models \sobo and \fobt indicates that the transition from a magnetically dominated to a rotationally dominated configuration occurs near $\Le\approx 1$, i.e.\ when the \alfven timescale is comparable to the rotation period. 

\begin{figure*}
    \centering
    \includegraphics[width=\textwidth]{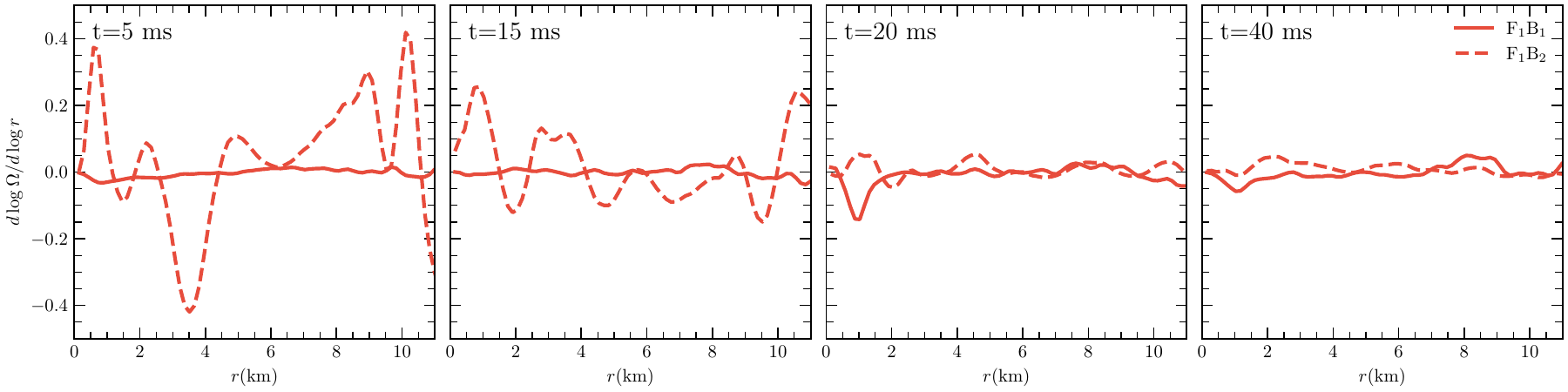}
    \caption{Evolution of the azimuthally averaged shear parameter
      in the equatorial plane for $\rm F_1$ models. Differential
      rotation develops faster and stronger in the
      magnetically dominated case \fobt.}
     \label{fig:qpar}
\end{figure*}

\begin{figure}
    \centering
    \includegraphics[width=0.5\columnwidth]{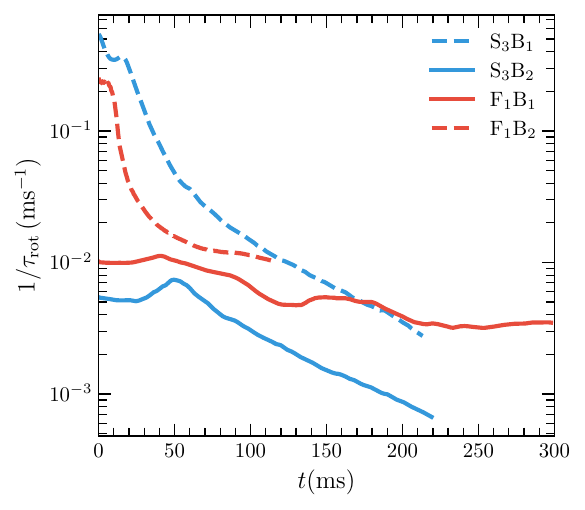}
    \caption{Evolution of the inverse of the rotational timescale
    ($1/\tau_{\rm rot}$) for representative models. Rotation-dominated
    runs (\sthbt, \ftbo), shown with solid lines, maintain values an order of magnitude lower than their magnetically dominated counterparts plotted with dashed lines (\sthbo, \ftbt), allowing sustained toroidal-field growth due to a weak damping term.}
    \label{fig:damp}
\end{figure}

A key mechanism identified in our simulations is that differential rotation develops at early times in our initially rigidly rotating systems, qualitatively similar to~\cite{Pinas:2025bpq}, and affects the field evolution on longer timescales simulated here. This is illustrated in Fig.~\ref{fig:qpar}, which shows the evolution of the shear parameter $q = d\log\Omega/d\log r$, i.e.\ the log-log derivative of the angular velocity with respect to the coordinate radius. Differential rotation develops faster for larger initial magnetic field strength but later subsides, even in fast rotators, as \alfven time increases. Given the presence of a strong initial poloidal field $B_p$, differential rotation drives a $\Omega$-dynamo, and the toroidal field $B_\phi$ evolves according to \cite{Barrere:2022A&A...668A..79B, Eksi:2026arXiv260321103E}:
\begin{equation}
    \frac{dB_\phi}{dt}=q\Omega B_p-\frac{B_\phi}{\tau_{\rm{rot}}},
    \label{eq:omegaeff}
\end{equation}
where $\Omega$ is the stellar angular velocity and $\tau_{\rm{rot}}=\tau_{\rm A}^2/P=P/\Le^2$ is the rotation-modified instability timescale (which differs from the initial period $P$).
$\Le<1$, i.e. the rapid rotation regime, therefore implies not only larger $\Omega$, thus strengthening the forcing term, but also a longer damping timescale. This is illustrated in Fig.~\ref{fig:damp}, where the time evolution of $1/\tau_{\rm rot}$ is one order of magnitude lower in the rotation-dominated models \sthbt and \fobo compared to their magnetically dominated counterparts \sthbo and \fobt.
The toroidal field can therefore grow persistently, while the preferred sense of
rotation selects a dominant handedness for the winding. The growth of toroidal field builds up a net linkage between the poloidal and toroidal fields that is, a finite magnetic helicity. In astrophysical NSs, this
winding process is expected to reach completion on timescales of order
$\tau \approx 10$--$100\,\tau_A$, at which point plasmoid instabilities
drive reconnection in the equatorial regions, allowing the field to
relax to a helical equilibrium~\cite{Stewart:2022, Loureiro16}. Once
a finite helicity is established, the configuration becomes
topologically protected and further large-scale field
decay is halted~\cite{Taylor:1974PhRvL..33.1139T,
Taylor:1986RvMP...58..741T}.
\begin{figure}[t]
    \centering

    \begin{minipage}{0.48\columnwidth}
        \centering
        \includegraphics[width=\linewidth]{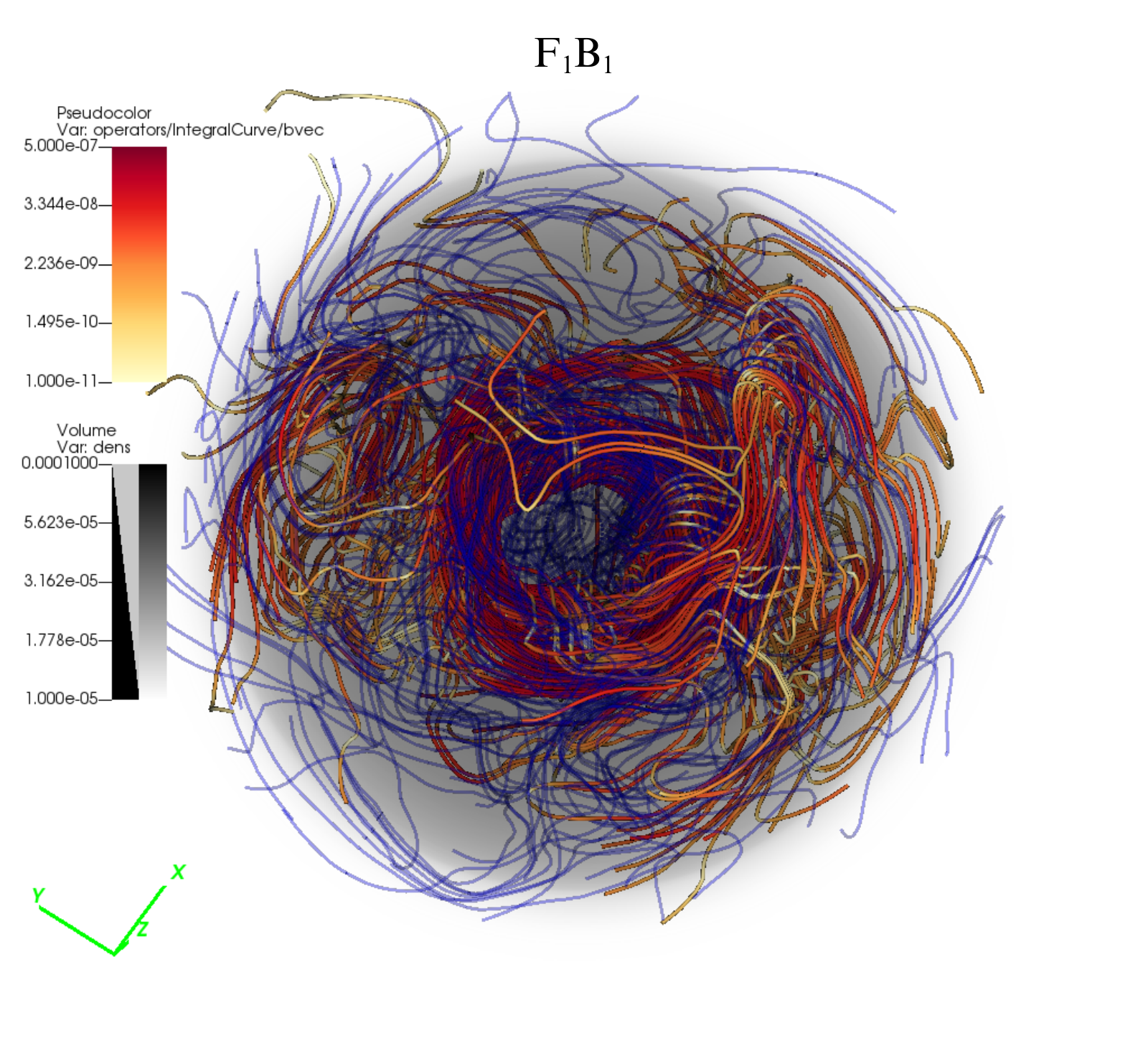}
    \end{minipage}
    \hfill
    \begin{minipage}{0.48\columnwidth}
        \centering
        \includegraphics[width=\linewidth]{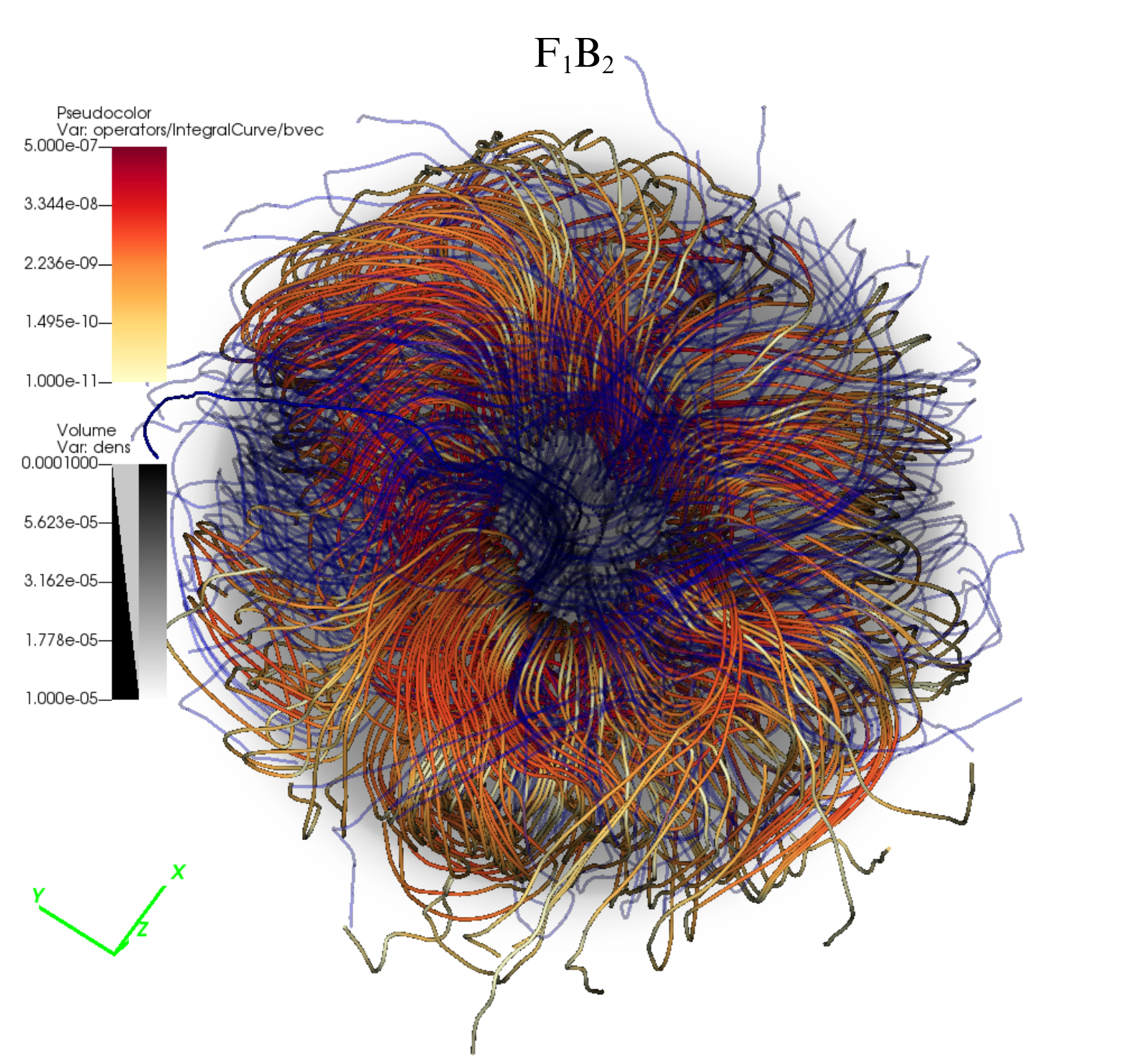}
    \end{minipage}

    \caption{Three-dimensional magnetic field-line structure plotted at cumulative \alfven time $T_A\sim 15$ for the rotation-dominated model \fobo ($\Le < 1$; left)
and the magnetically dominated model \fobt ($\Le > 1$; right). Orange-to-red field lines are seeded at a radius of $4.5$\,km and colour-coded by $B_\phi^2$ (orange: low, red:
high); blue lines are seeded near the stellar centre and trace the
poloidal field structure. In \fobo,
differential winding organises the field into a coherent twisted-torus
configuration with a dominant toroidal component concentrated in the
interior. In \fobt, the toroidal field is relatively weak and magnetic field lines show predominantly poloidal structures. The contrast illustrates the qualitatively
different end states of the two regimes identified in this work.}
    \label{fig:3dstream}
\end{figure}

Figure~\ref{fig:3dstream} confirms the qualitative difference between the two regimes, plotted at $T_A \sim 15$, where blue lines trace
the poloidal field seeded near the stellar core and orange-to-red
lines seeded at $r = 4.5$\,km are colour-coded by $B_\phi^2$. In \fobt the field lines show dominantly poloidal structure, following ordered meridional circuits, with only weak toroidal winding, while in \fobo differential winding produces a tangled, toroidally dominated, equatorially
concentrated structure.

To explore the implications of our results for the observable NS population, we extrapolate to a generic pulsar characterised by rotation period $P$ and surface magnetic field $B$. The transition between magnetically dominated and rotationally dominated configurations occurs at $\Le\approx1$. Therefore, we assume that the mean field $B$ computed in our simulations can serve as a proxy for the dipole field inferred from pulsar spin-down. We note that this is a significant approximation as our simulations do not include an exterior dipolar component. Taking the spin-down inferred field for the rotating vacuum dipole model, $B \approx 3.2 \times 10^{19} \sqrt{P\dot{P}}$\,G \cite{Shapiro:1983du} (with $P$ in seconds), we can delineate the boundary between magnetically and rotationally dominated magnetised NSs.

\begin{figure}[t]
    \centering
    \includegraphics[width=0.75\textwidth]{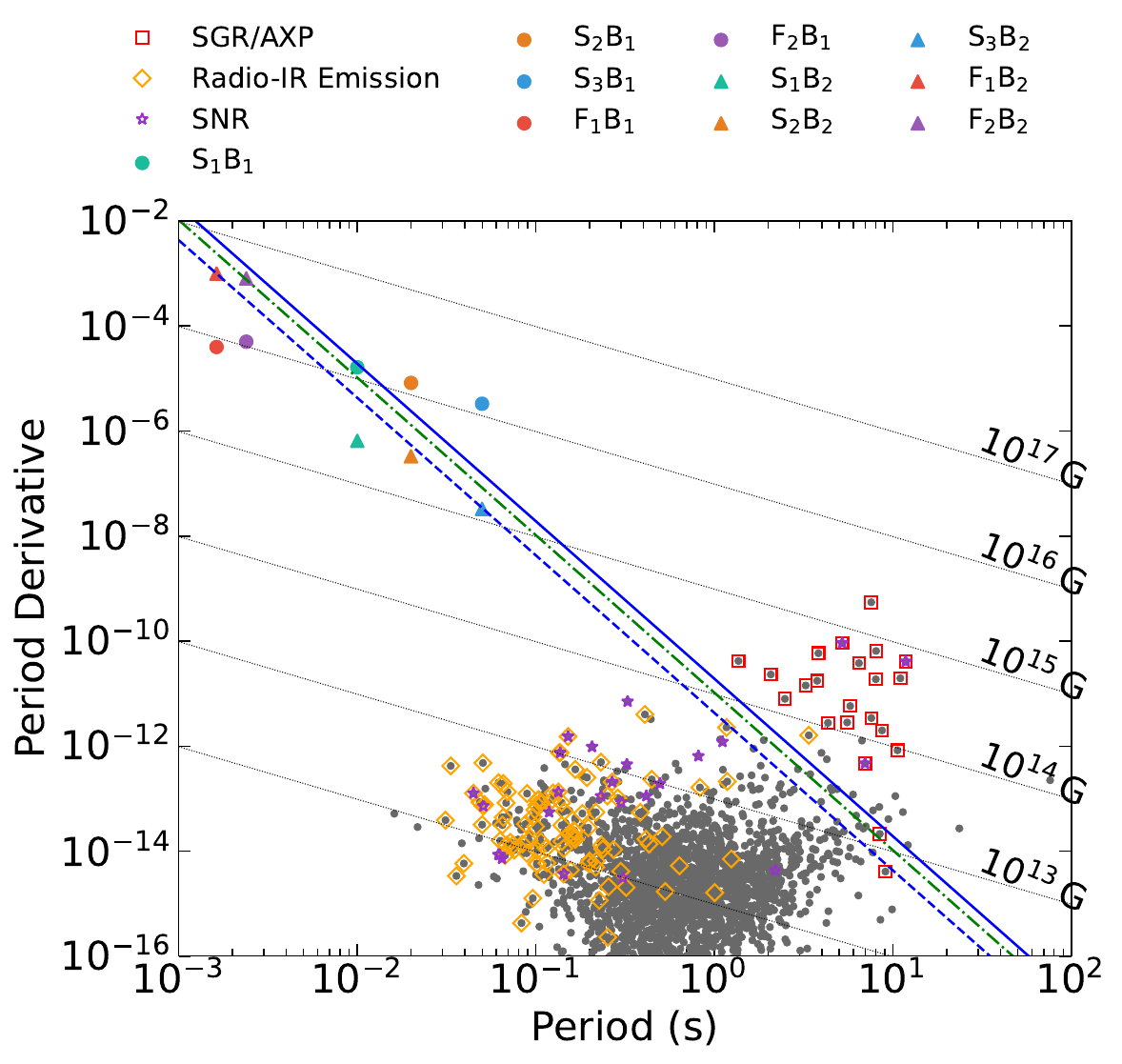}
    \caption{$P$--$\dot{P}$ diagram showing the boundary at which the topology of the magnetic field switches from a poloidal, `magnetically' dominated structure, to a twisted torus `rotationally' dominated configuration. Solid and dashed lines are derived from models \sobo and \ftbo respectively; the green line uses the average $\alpha_{\rm P}$. Magnetars are shown as red boxes; pulsars associated with supernova remnants as magenta stars; pulsars observed in the infrared are shown in yellow diamonds. Our simulations are indicated by coloured dots for sequence $\rm B_1$ and triangles for $\rm B_2$ sequences.}
    \label{fig:Ppdot}
\end{figure}

Figure~\ref{fig:Ppdot} shows this boundary overlaid on the $P$--$\dot{P}$ diagram, produced using \texttt{python} interface \texttt{psrqpy} \cite{psrqpy} using data from the ATNF pulsar catalogue \cite{Manchester:2004bp} for all pulsars with measured $P$ and $\dot{P}$. The solid line is derived from model \sobo and the dashed line from \ftbo (see Supplementary Material for details); these lines separate NSs into two regions, with systems above the line being magnetically dominated ($\Le>1$) and those below being rotationally dominated ($\Le<1$). Magnetars are marked with red boxes and pulsars associated with supernova remnants with magenta stars, radio pulsars also observed in the infrared are shown in yellow diamonds; our simulations are indicated with coloured dots and triangles.\\

Strikingly, nearly all systems lying above the boundary and therefore in the magnetically dominated regime, are magnetars, for which one would consequently expect a predominantly poloidal internal field. In this case, the continuous field decay and active instabilities observed in our simulations provide a natural physical mechanism for the bursting activity of Soft Gamma Repeaters and Anomalous X-ray Pulsars. We note that two systems identified as magnetars in Fig.~\ref{fig:Ppdot}, namely SGR~J1822$-$1606 and SGR~0418+5729, lie close to, rather than clearly above, the boundary separating the two regimes. These are the so-called `low-field' magnetars, for which the relatively low dipole field inferred from spin-down cannot account for the observed bursting behaviour or X-ray spectra, and for which a hidden strong internal toroidal component has been invoked \cite{Guver:2011a, Tiengo:2013, Perna:2013}.\\

Our analysis provides a natural explanation for the presence of strong internal toroidal fields in these low-field magnetars. Our simulations indicate that when the initial \alfven timescale exceeds the rotation period in a newly born NS, differential rotation can wind up a strong internal toroidal component even before the star settles into a slowly rotating state.
Furthermore, magnetothermal simulations of crustal magnetic configurations have suggested that bursting activity is linked to a strong toroidal component \cite{Igoshev:2021, Igoshev:2025NatAs...9..541I}. Our results are consistent with this picture: a strong toroidal field wound up by differential rotation at birth will, once the differential rotation has decayed, become unstable and evolve toward the magnetically dominated configurations we have calculated, thereby providing a physical trigger for the observed bursting behaviour. Differences between individual sources may therefore reflect differences in progenitor properties and the degree of differential rotation at birth ---for instance, between stars born in core-collapse supernovae and those formed in binary NS mergers.

For the bulk of the standard pulsar population, on the other hand, we expect mixed configurations in which a strong toroidal component develops alongside the poloidal field. It is therefore plausible that most pulsars harbour an internal toroidal component up to an order of magnitude stronger than the exterior dipole inferred from spin-down.
This has important consequences for gravitational wave emission: magnetic fields deform the star and produce a time-varying mass quadrupole as it rotates, generating quasi-monochromatic, long-lived continuous gravitational wave signals. For young, non-superconducting stars the intrinsic signal amplitude scales quadratically with the mean field strength \cite{Haskell:2007bh}, and linearly for older, colder stars with superconducting interiors \cite{Lander:2012a}. A significantly stronger interior field could therefore produce a stronger signal than the inferred exterior dipole alone would suggest, making a number of young pulsars potentially interesting targets for next-generation gravitational wave detectors such as the Einstein Telescope and Cosmic Explorer \cite{Bluebook:2026JCAP...03..081A}. Transitional NS sources evolving from a magnetically dominated poloidal configuration toward a rotationally dominated strong toroidal field, i.e.\ the low-field magnetars discussed above, could be promising targets for space-based detectors sensitive to the sub-20\,Hz band, such as DECIGO and BBO \cite{Pagliaro:2025MNRAS.540.1006P}.

\section{Supplementary Material}
\label{sec:methods}

\subsection{Simulations}

Simulations are performed in GR by evolving initially uniformly rotating NS with interior poloidal magnetic fields of different field strengths. 
Initial data for the simulations is generated using \texttt{RNS} code \cite{Stergioulas:1994ea}. The equilibrium properties of these models, namely central rest mass density $\rho_c$, gravitational mass $M$, equatorial coordinate radius $r_e$, ratio of polar to equatorial radius $r_p/r_e$, 
and rotation period $P$ are summarized in Table \ref{tab:models}. For all models, we adopt a polytropic equation of state with $N = 1$. These models belong to the AU sequence~\cite{Dimmelmeier:2005zk}, constructed for a fixed rest mass of $M_0 = 1.506\,M_\odot$.
 
The magnetic field is initialized by a purely toroidal vector potential of the form,     
\begin{equation}
\label{eq:vcpt}
A_{\phi}=A_b\,{\rm max}\left(p-0.04\,p_{\rm max},0.0\right),
\end{equation}
which generates a purely poloidal magnetic field which is completely confined within the star. The parameter $A_b$ controls the strength of the magnetic field. The resulting field strength is determined by the pressure distribution within the star. The simulations labeled ${\rm  F}$ correspond to fast rotators, while those labeled ${\rm S}$ represent slow rotators. We adopt the $\mathrm{F_2}$ and $\mathrm{S_1}$ models as control cases to determine the threshold magnetic-field strength at which the stellar rotation period begins to have a dynamical impact on the evolution of magnetic field.
For runs labeled $\rm B_1$, the canonical magnetic-field parameter is fixed at $A_b = 5$ across all models. The $\rm B_2$ series is constructed by scaling the magnetic field
strength of each $\rm B_1$ model to shift it across the $\Le = 1$ boundary, allowing us to isolate the effect of field strength on the
regime transition. The average initial magnetic-field strengths for these models are listed in Table~\ref{tab:models}.

Numerical-relativity simulations are performed using the code \AK \citep{Stone:2024,Zhu:2024utz,Fields:2024pob} which solves the equations of ideal GRMHD in dynamical spacetime. The spacetime evolution equations are solved as the Z4c free-evolution scheme of 3+1 Einstein equations \cite{Bernuzzi:2009ex,Hilditch:2012fp}; GRMHD equations are solved in Eulerian conservative form using conservative finite-volume methods. The magnetic field is evolved using upwind constrained transport scheme~\cite{Gardiner:2007nc}. Reconstruction of primitive variables at cell interfaces is performed with the {PPMX} scheme \cite{Colella:2008}; numerical fluxes are calculated using Harten--Lax--van Leer--Einfeldt (HLLE) approximate Riemann solver augmented with first-order flux correction \cite{Fields:2024pob}. The time evolution is performed using a 3rd order strong-stability preserving Runge-Kutta scheme \cite{Gottlieb:2009a}. We use a static, block-structured refined grid. The finest refinement level spans a cubic domain from $-60$km to $+60$km in each direction, with a spatial resolution of $\Delta x=230$m.

\begin{table}[t]
    \centering
    \begin{tabular}{lcccccccccc}
            \toprule
        Model & $\rho_c(\times 10^{-3})$ & $M$ &  $r_e$ & $r_p/r_e$ & $P$ (ms) & $B_1\,{\rm \times10^{15} G}$ & $B_2\,{\rm \times10^{15} G}$ & \multicolumn{2}{c}{$\Le$} \\         \cmidrule(lr){9-10}
        & & & & & & & 
        & $B_1$ & $B_2$ \\
        \midrule 
        $\rm F_1$ & 0.978 & 1.411  & 10.06 & 0.7800  & 1.63 & 8.15 & 40.77 &0.12 &0.64\\        $\rm F_2$ & 1.165 & 1.404  & 8.71 & 0.9190  & 2.39 & 11.09 & 44.39 &0.23 &0.91\\        \hline
        $\rm S_1$ &  1.275  & 1.4004  &  8.15 & 0.9965  & 10.84 & 12.94 & 2.58 & 0.88 &0.20\\
        $\rm S_2$ &  1.278  & 1.4001  &  8.13 & 0.9990  & 20.52 & 13.00 & 2.60 & 2.08 &0.41 \\
        $\rm S_3$ &  1.279  & 1.4000  &  8.127 & 0.9998 & 50.00 & 13.02 & 1.30 & 5.22 &0.52\\
        \bottomrule
    \end{tabular}
    \caption{Equilibrium properties of the neutron star models used in
this work: central rest-mass density $\rho_c$, gravitational mass $M$,
equatorial coordinate radius $r_e$, polar-to-equatorial radius ratio
$r_p/r_e$, rotation period $P$, initial average magnetic
field strengths $B_1$ and $B_2$, and the corresponding Lehnert numbers
$\Le \equiv P/\tau_A$. All quantities except $P$
and $B_{1,2}$ are in geometric units with solar masses ($c=G=M_{\odot}=1$).}
    \label{tab:models}
\end{table}

\subsection{Diagnostics}

We monitor the total magnetic energy ($E_{\rm B}$), energy stored in toroidal ($E_{\rm {tor}}$) and poloidal ($E_{\rm {pol}}$) components as
\begin{equation}
E_{\rm B}=\frac{1}{2}\int B_{i}B^{i}\sqrt{\gamma}dx^3,   
\quad
E_{\rm {tor}}=\frac{1}{2}\int B_{\phi}B^{\phi}\sqrt{\gamma}dx^3,   
\quad
E_{\rm {pol}}=E_{\rm B}-E_{\rm {tor}},   
\label{eq:pole}
\end{equation}
where $\gamma \equiv \det(\gamma_{ij})$ is the determinant of the
spatial metric, and $B^i$ are the magnetic field components measured
by the Eulerian observer.
The magnetic field evolves on a characteristic timescale, referred to as the Alfvén crossing timescale, which is determined as:
\begin{equation}
\tau_{\rm A}=\frac{2R_{\rm NS}\sqrt{<\rho>}}{<B>},    
\end{equation}
where $<\rho>$ and $<B>$ are the volume average density and magnetic field strength evaluated within the bulk of the star, and $R_{\rm NS}$ is the polar radius of the star.
As these quantities evolve throughout the simulation, we define the cumulative \alfven crossing period as follows, following the definition in \cite{Sur:2020hwn}:
\begin{equation}
\label{eq:t_cum}
T_A=\int_0^t\frac{dt}{\tau_{\rm A}(t)}.
\end{equation}
To account for the stabilizing influence of rapid rotation on magnetic instabilities, we introduce the rotationally modified Alfvén timescale \(T_r\). In rapidly rotating systems, the Coriolis force suppresses the growth rate of non-axisymmetric MHD instabilities, causing them to evolve on a timescale longer than the ordinary Alfvén time \(\tau_{\rm A}\). Pitts \& Tayler \citep{Pitts:1985} showed that the effective instability timescale in the limit \(P \ll \tau_{\rm A}\) scales as $\tau_{\rm rot} \sim \frac{\tau_{\rm A}^{\,2}}{P}=\frac{\tau_{\rm A}}{\Le}$, where \(P\) is the stellar rotation period and we have defined the Lehnert number $\Le={P}/{\tau_{\rm A}}$. Following this, we define the cumulative rotational Alfvén crossing time \(T_r(t)\) as
\begin{equation}
T_r = \int_0^{t} \frac{dt}{\tau_{\rm A}(t)\sigma_{\rm r}},
\end{equation}
where $\sigma_{\rm r}=\tau_{\rm A}/P$ for the case $\tau_{\rm A}>P$, otherwise $\sigma_{\rm r}=1$. $P$ is the rotation period of the star. $T_r$ measures the elapsed time in units of the rotation-modified instability timescale rather than in ordinary Alfvén units. This rescaling provides a natural temporal normalization when comparing the onset of magnetic-field evolution across models with different rotation rates. 
In order to investigate the role of stellar rotation in structuring the magnetic field, we examine the time evolution of the magnetic helicity, which is defined as
\begin{equation}
H=\int \mathbf{A}\cdot\mathbf{B}\,dV,
\label{eq:helicity}
\end{equation}
where the volume integral is taken over the domain covered by the finest refinement level which completely encloses the star. 
Following \cite{Ciolfi:2012en}, we normalize helicity by a reference value $\tilde{H}=R_{\rm NS}E_{\rm EM}(0)/2$, corresponding to a configuration in which the toroidal and poloidal field components contribute equally, where $E_{\rm EM}$ is the electromagnetic energy calculated as 
\begin{equation}
E_{\rm EM}=\int \left[b^2 \left(W^2-\frac{1}{2}\right) -(\alpha b^0)^2\right]\sqrt{\gamma}dx^3,   
\label{eq:magen}
\end{equation}
where $W$ and $\alpha$ are the Lorentz factor and lapse function
respectively, and $b^\mu$ is the magnetic four-vector measured by the comoving observer
with invariant $b^2 \equiv b_\mu b^\mu$. \\
As \AK evolves $\mathbf{B}$, for the calculation of helicity we
reconstruct $\mathbf{A}$ from $\mathbf{B}$ using the cell-by-cell algorithm of \cite{Silberman:2018ioy}.

\subsection{Helicity evolution}

\begin{figure*}[t]
    \centering
    \includegraphics[width=0.95\textwidth]{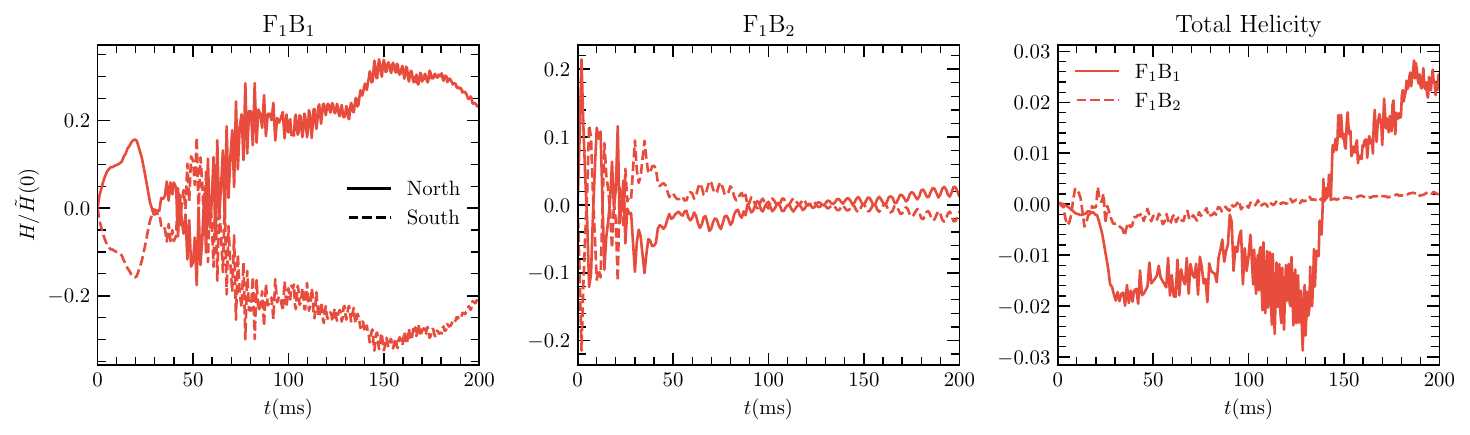}
    \includegraphics[width=0.95\textwidth]{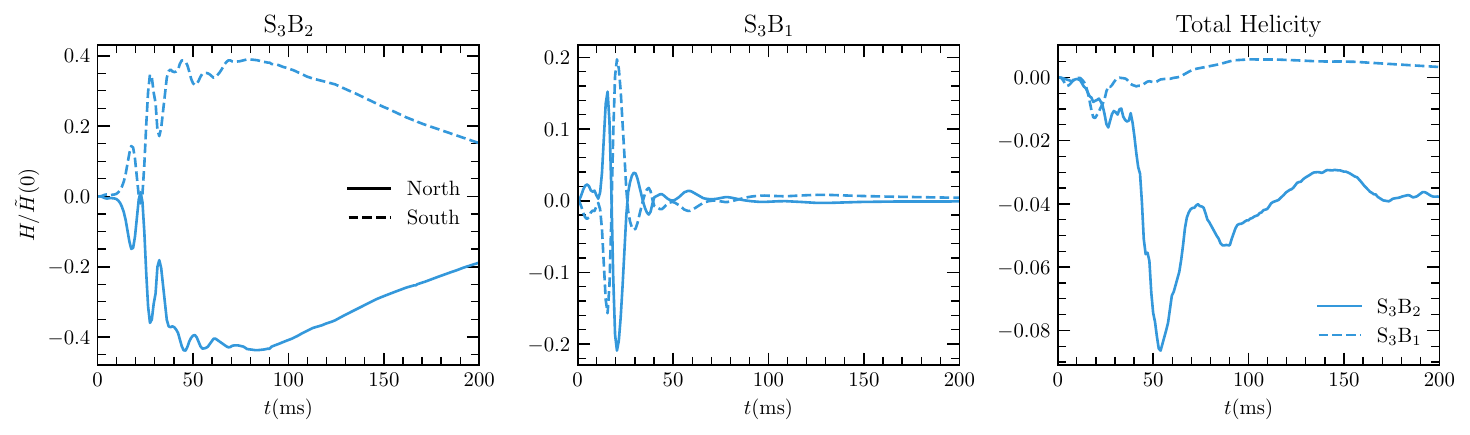}
    \caption{Magnetic helicity evolution for the fastest rotating
($P = 1.63$\,ms; top) and slowest rotating
($P = 50$\,ms; bottom) models. Left and middle
columns show the helicity accumulated in the northern and southern
hemispheres separately for the rotationally and magnetically dominated
models respectively. The right column shows the total helicity.}
    \label{fig:helicity}
\end{figure*}

Figure~\ref{fig:helicity} shows the evolution of magnetic helicity for the fastest  
($\rm F_1$) and slowest ($\rm S_3$) model. Since the initial configuration
is purely poloidal, the field lines begin with no twist or linkage resulting in zero helicity, as
expected for a purely poloidal field.
In the rotation-dominated case of these models \fobo and \sthbt,
the progressive winding builds helicity of opposite sign in the two
hemispheres as seen from the left column of Figure~\ref{fig:helicity}. The onset of kink instabilities then triggers a rapid rise in helicity. The later phases of nonlinear reorganization result in a preferred handedness, leading to a net helicity with a consistent sign, accompanied by strong subsequent oscillations
in total helicity for model \fobo. Once a net helicity is
established, the configuration becomes topologically protected, halting
further decay. In contrast, the magnetically dominated
models \fobt and \sthbo show rapid poloidal field decay with only a weak toroidal
component developing before the field relaxes toward a quasi-equilibrium.
The presence of only a transient helicity signal that quickly damps, confirms that
the winding is not sustained by the underlying dynamics.

\subsection{Extrapolating to the pulsar population}
The transition between magnetically and rotationally dominated regimes
occurs when the \alfven crossing timescale equals the rotation period. In order to extend our analysis to map this boundary onto the observed NS population,
we express this condition as
\begin{equation}
  \frac{\alpha_{\rm P}}{B}=P
\end{equation}
with $P$ the rotation period and $\alpha_{\rm P}$ a constant, such that $\tau_{\rm A}=\alpha_{\rm P}/<B>$. This approximation allows us to obtain the value of the constant $\alpha_{\rm P}$ from our results. We take an average value of $\alpha_{\rm P} = 1.03 \times 10^{14}$ from models \sobo and \fobt. Using this value, we estimate that the average magnetic-field strength corresponding to the rotation period of $\rm F_2$ is $B = 4.3 \times 10^{16}\,\mathrm{G}$. Consequently, any initial magnetic-field strength greater than $4.3 \times 10^{16}\,\mathrm{G}$ is expected to lead to a magnetically dominated evolution for $\rm F_2$. 

This inference is confirmed by the \ftbt simulation, for which the average initial magnetic-field strength is $4.4 \times 10^{16}\,\mathrm{G}$, marginally exceeding the estimated threshold. The energies stored in the toroidal and poloidal components are shown by the purple curves in the bottom middle panel of Fig.~\ref{fig:energy_b1}. The evolution of this model is magnetically dominated, characterised by stronger magnetic-field losses and a comparatively weaker toroidal-field component.
Similarly for slowly rotating models $\rm S_2~and~S_3$, the derived values of the magnetic field, that separate magnetically and rotation dominated evolution are $5.15\times10^{15}\,G$ and $2.06\times10^{15}\,G$ respectively. The magnetic field strengths taken for models \stbo and \sthbo (see Table \ref{tab:models}) lie in the magnetically dominated regime, while in \stbt and \sthbt field strengths correspond to the rotation dominated regime. The energy evolution of \sthbt shown in Fig.~\ref{fig:energy_b1} shows that initially $E_{\rm tor}$ is weaker in comparison to other runs because the source term $q\Omega$ (Eq. \ref{eq:omegaeff}) is small due to slower rotation in this model, however, the damping term is also weak, which allows for a continuous growth of the toroidal component which reaches up to 80\% of total energy by the end of simulation.

\begin{acknowledgments}
  We thank Aurora Capobianco for discussions.
  RJ, SB, BH acknowledge support for the MERLIN project. MERLIN is
  funded by the Deutsche Forschungsgemeinschaft (DFG) and the Narodowe
  Centrum Nauki (NCN) OPUS-LAP grant number 2022/47/I/ST9/01494, under
  the EU weave initiative.
  SB acknowledges support by the DFG project ``Magnetfelddynamik in
  Neutronensternen Sternen'' MERLIN (MERLIN; BE 6301/6-1
  Projektnummer: 524726453).
  SB acknowledges support by the EU Horizon under ERC Consolidator
  Grant, no. InspiReM-101043372.
  This article is based upon work from COST Action SCALES, CA24139, supported by COST (European Cooperation in Science and Technology)
  
  We acknowledge the EuroHPC Joint Undertaking for awarding this
  project access to the EuroHPC supercomputers, LEONARDO, hosted by
  CINECA (Italy) and the LEONARDO consortium; LUMI, hosted by CSC
  (Finland) and the LUMI consortium; and KAROLINA hosted by
  IT4Innovations National Supercomputing Center (Czech Republic)
  through an EuroHPC Benchmark Access call (EHPC-BEN-2024B10-018). 
  We acknowledge ISCRA for awarding this project access to the LEONARDO supercomputer, owned by the EuroHPC Joint Undertaking, hosted by CINECA (Italy) through projects IsCc8\_MERLIN0 and IsB31\_MERLIN1.
  The authors gratefully acknowledge the Gauss Centre for Supercomputing e.V. (www.gauss-centre.eu) for funding this project by providing computing time on the GCS Supercomputer JUWELS \citep{JUWELS:2021a} at Jülich Supercomputing Centre (JSC) through project MAGWAVE.
  The authors gratefully acknowledge the Gauss Centre for Supercomputing e.V. (www.gauss-centre.eu) for funding this project by providing computing time on the GCS Supercomputer SuperMUC-NG Phase 2 at Leibniz Supercomputing Centre (www.lrz.de) through projects pn67xo and pn76li.

Computation and analysis for this article were also performed using computer cluster at the Nicolaus Copernicus Astronomical Center of the Polish Academy of Sciences (CAMK PAN).

  Computations were also performed on the ARA and Draco clusters at
  Friedrich Schiller University Jena and on the {\tt Tullio} INFN
  cluster at INFN Turin. The ARA cluster is funded in part by DFG
  grants INST 275/334-1 FUGG and INST 275/363-1 FUGG, and ERC Starting
  Grant, grant agreement no. BinGraSp-714626.
\end{acknowledgments}

\end{document}